\documentclass[a4paper, 12pt]{article}

\usepackage{amssymb}
\usepackage{amsthm}
\usepackage{amsmath,amsfonts}
\usepackage{algorithm}
\usepackage{algorithmic}
\usepackage{graphicx}
\usepackage{epstopdf}
\usepackage{thmtools,thm-restate}
\usepackage{lineno,hyperref}
\usepackage{color}
\usepackage{multirow}
\usepackage{booktabs}
\usepackage{bigstrut}
\usepackage{hyperref}
\usepackage[table]{xcolor}
\usepackage{tabularx}
\usepackage{array}
\usepackage{enumitem}
\usepackage{authblk}

\newtheorem{thm}{Theorem}
\newtheorem{lem}[thm]{Lemma}
\newtheorem{mydefinition}{Definition}

\title{Online Scheduling of Time-Critical Tasks to Minimize the Number of Calibrations}

\author[1,2]{Zuzhi Chen}
\author[1,2]{Jialin Zhang \thanks{Corresponding author: zhangjialin@ict.ac.cn}}

\affil[1]{Institute of Computing Technology, Chinese Academy of Sciences, Beijing, China}
\affil[2]{University of Chinese Academy of Sciences, Beijing, China}
\date{}

\begin{document}

\maketitle

\begin{abstract}
We study the online scheduling problem where the machines need to be calibrated before processing any jobs.  To calibrate a machine, it will take $\lambda$ time steps as the activation time, and then the machine will remain calibrated status for $T$ time steps. The job can only be processed by the machine that is in calibrated status. Given a set of jobs arriving online, each of the jobs is characterized by a release time, a processing time, and a deadline. We assume that there is an infinite number of machines for usage. The objective is to minimize the total number of calibrations while feasibly scheduling all jobs.
                           
For the case that all jobs have unit processing times, we propose an $\mathcal{O}(\lambda)$-competitive algorithm, which is asymptotically optimal. When $\lambda=0$, the problem is degraded to rent minimization, where our algorithm achieves a competitive ratio of $3e+7(\approx 15.16)$ which improves upon the previous results for such problems.
\end{abstract}

\section{Introduction}
\label{Sec:Intro}
    Online scheduling captures the \emph{online} nature in many real-life scheduling scenarios. The jobs arrive one by one and we need to schedule the arriving jobs before knowing the complete information about the entire problem instance. Numerous works discuss various problem settings including machine models, processing formats, and objective functions corresponding to rich practical tasks \cite{DBLP:reference/crc/PruhsST04,DBLP:books/tf/09/Albers09}.
        
    In this paper, we consider the scheduling with calibrations which was first introduced by Bender et al. \cite{DBLP:conf/spaa/BenderBLMP13}. It was motivated by a nuclear weapon test program called Integrated Stockpile Evaluation (ISE). The original setting is that the machine needs to be calibrated before it can process jobs. After one calibration, the machine stays in calibrated status for $T$ time steps. The idea of calibration scheduling can be extended to a wide variety of industrial applications which involve high-precision measurement, high-quality products, or extra safety requirements. Machines that are used under these situations may require calibrations. For example, in the pharmaceutical and food processing industry, the instruments for manufacturing must be calibrated to meet certain standards under regulations \cite{us2003current, ng2009good}, for the purpose of quality or environmental management. In practical scenarios, machine calibration cannot be done instantaneously. For example, in industrial applications, the disinfection of a machine can be viewed as the calibrating period of a machine, and this will require a non-negligible time to finish. 
        
    With the above context, in this paper, we focus on the following online scheduling problem: machine calibration will take $\lambda\geq 0$ time steps, then the calibrated status can last for $T$ time steps. The machine can process jobs only when it is in the calibrated status. Each job is characterized by a release time, a processing time, and a deadline. The jobs are arriving online, i.e., the algorithm only learns the job when the job is released. We assume that there is an infinite number of machines. The objective is to minimize the total number of calibrations while ensuring that all jobs are feasibly scheduled.
        
    The assumption of an infinite number of machines may seem impractical, but we keep such an assumption for the following three reasons. First, if the cost of calibration is high, in the long-term perspective, the cost of purchasing machines is far less than the cost of calibrations.  Second, in some applications like cloud rental, where the calibration itself can be viewed as a service, the infinite machines assumption is equivalent to the assumption that the market can provide enough services for the customers. In this case, the assumption is quite plausible. Third, if the number of machines is limited, even we know all jobs in advance (offline case), it is still NP-hard to decide whether there exists a feasible schedule for all of the jobs \cite{Brucker_2004}. In a special case where the processing time of any job is 1 (called unit processing time case), feasibility in the offline case can be easily checked. But if we want to make sure of the feasibility in the online case with a limited number of machines, the only thing we can do is to immediately schedule any released job. Thus, if we expect any non-trivial online algorithm with the guaranteed performance, we must make a trade-off between the number of calibrations and some other factors, for example, the number of successfully scheduled jobs or the total delay time of the jobs. In this paper, we would like to focus on the number of calibrations, and the setting of infinite machines can naturally avoid this infeasibility problem.
        
    Under the online setting, no online algorithm can always obtain the optimal solution with only incomplete information. Thus, we adopt the notion of \textbf{\textit{competitive ratio}} for the measure of the online algorithm. Given an instance $Ins$, let $ALG(Ins)$ be the number of calibrations used by the algorithm \textbf{ALG} under instance $Ins$, and let $OPT(Ins)$ be the optimum number. We say the algorithm \textbf{ALG} is $c$-competitive if there exists a constant number $a$, and for any instance $Ins$, we have $ALG(Ins)\leq c\cdot OPT(Ins)+a$.

    \subsection{Our Contributions}
        
    We consider the case that all jobs have unit processing times and let the integer $\lambda\geq 0$ be the activation time of a machine, i.e., the time units required for calibrating a machine. 
        
    For the special case that $\lambda=0$, our problem is degraded to rent minimization problem where all the jobs have unit processing times. We propose an algorithm that achieves a competitive ratio of $3e+7(\approx 15.16)$, which improves upon the previous results(at least $30$) of the rent minimization problem.
        
    For any $\lambda \geq 0$, our main result is an online algorithm with a competitive ratio of $3(e+1)\lambda+3e+7$. We also give a lower bound of $\max\left\lbrace e,\lambda\right\rbrace$ for any online algorithms. This implies our algorithm is asymptotically optimal for the order of $\lambda$. Besides, in our understanding, the previous results on rent minimization cannot directly be extended to the case of $\lambda >0$.
        
    \subsection{Related Work}
    The study of scheduling with calibrations was initiated by Bender et al. \cite{DBLP:conf/spaa/BenderBLMP13}, where they considered the offline scheduling problem. Bender et al. \cite{DBLP:conf/spaa/BenderBLMP13} gave an optimal solution for a single machine case and a $2$-approximation algorithm for multiple machines case. Later, Chen et al.\cite{DBLP:conf/spaa/ChenLL019} proposed a PTAS for this problem via dynamic programming. However, the hardness result for this problem in the offline version is still open.
        
    Then, Fineman and Sheridan \cite{DBLP:conf/spaa/FinemanS15} extended the study of the above offline calibration scheduling problem to the arbitrary processing time case. They assumed $m$ machines are feasible to schedule all jobs and proved that an $s$-speed $\alpha$-approximation algorithm for the machine minimization problem implies an $s$-speed $\mathcal{O}(\alpha)$-machine $\mathcal{O}(\alpha)$-approximation algorithm for the calibration scheduling problem with arbitrary processing time jobs.
        
    As for the scenario of online scheduling with calibrations, it was first introduced by Chau et al. \cite{DBLP:conf/spaa/ChauLMW17}. They removed the deadline requirement and considered the linear combination of total weighted flow time and the cost of calibrations as the objective function. In this way, they avoided the difficulty we mention above in guaranteeing the feasibility for the job with a deadline if the machines are limited. They gave several online algorithms with a constant competitive ratio for this problem.
        
    All the above work assumed that calibrating a machine can be done instantaneously. Angel et al. \cite{DBLP:conf/faw/AngelBCZ17} was the first to consider the scenario that calibrating a machine will require $\lambda$ time units. They gave a polynomial running time algorithm for the offline scheduling with unit processing time jobs.
        
    Another scheduling problem that is closely related to online scheduling with calibration is the rent minimization problem, which is proposed by Saha \cite{DBLP:conf/fsttcs/Saha13}. In rent minimization, each machine can only be used for $T$ time steps, and the objective is to minimize the total number of machines while feasibly scheduling all jobs. The job is also arriving online and is characterized by a release time, a processing time, and a deadline. Saha \cite{DBLP:conf/fsttcs/Saha13} gave an asymptotically optimal algorithm with $\mathcal{O}\left(\log \frac{p_{max}}{p_{min}}\right)$-competitive ratio (where $p_{max}$ and $p_{min}$ is the maximum and minimum processing time of the input jobs.). For the case of unit processing time jobs, Saha \cite{DBLP:conf/fsttcs/Saha13} gave an $\mathcal{O}(1)$-competitive algorithm without an explicit constant competitive ratio. However, following the analysis of \cite{DBLP:conf/fsttcs/Saha13}, the competitive ratio of their algorithm for the case of unit processing time should be at least $30$.
        
    In this paper, we focus on the case of unit processing time jobs. We summarize the previous main results of scheduling with calibrations and unit size jobs in table \ref{table_unit}.
        
    \begin{table}[htbp]
        \centering
        \caption{Summary: results of unit size jobs}
        \begin{tabular}{|c|c|c|c|}
            \toprule
             Input & \multicolumn{1}{p{4.19em}|}{Machines} & \multicolumn{1}{c|}{Results} & \multicolumn{1}{c|}{Paper} \\
            \midrule
             \multirow{4}[8]{*}{Offline} & \multicolumn{1}{c|}{\multirow{2}[4]{*}{Single}} & Optimal & Bender et al. \cite{DBLP:conf/spaa/BenderBLMP13} \\
            \cmidrule{3-4}    \multicolumn{1}{|c|}{} &       & optimal (with activation  time $\lambda$) & Angel et al. \cite{DBLP:conf/faw/AngelBCZ17} \\
            \cmidrule{2-4}    \multicolumn{1}{|c|}{} & \multicolumn{1}{c|}{\multirow{3}[6]{*}{Multi}} & 2-approximation & Bender et al. \cite{DBLP:conf/spaa/BenderBLMP13} \\
            \cmidrule{3-4}    \multicolumn{1}{|c|}{} &       & PTAS & Chen et al.\cite{DBLP:conf/spaa/ChenLL019} \\
            \cmidrule{1-1}\cmidrule{3-4}    Online &       & $O(1)$-competitive ratio (no deadlines) & Chau et al. \cite{DBLP:conf/spaa/ChauLMW17} \\
            \bottomrule
            \end{tabular}%
            \label{table_unit}%
        \end{table}%

    \subsection{Overview}
    We present some definitions and preliminary results in Section \ref{Sec:Preliminaries}. We divide the unit processing time jobs into two classes, $\alpha$-long window jobs and $\alpha$-short window jobs, where $\alpha\in\left(0,1\right)$, and design different algorithmic strategies for these two cases. We first propose two deterministic algorithms for the case if all jobs are $\alpha$-long window jobs (Section \ref{Sec:alpha_long}) and the case if all jobs are $\alpha$-short window jobs (Section \ref{Sec:alpha_short}). Then we combine the two algorithms for the general case and decide the optimal value of parameter $\alpha$ in Section \ref{Sec:Integrated Algorithm}. Finally, we summarize our results in Section \ref{Sec:Conclusion}.

\section{Preliminaries}
    \label{Sec:Preliminaries}
    \subsection{The Model}
    We call the problem that we proposed \textbf{\textit{online time-critical task scheduling with calibration}}. The formal definition of the problem is stated as follows:
        
    We assume that there is an infinite number of machines, and
    \begin{enumerate}
        \item Calibrating a machine will take $\lambda$ time steps as the \textbf{\textit{activation time}}. Then the machine will remain in the \textbf{\textit{calibrated status}} for $T$ consecutive time steps.
        \item The machine can process jobs only when it is in the calibrated status.
        \item One machine can only process at most one job at any given time.
    \end{enumerate}
        
    Let $\mathcal{J}=\left\lbrace j_1,\dots,j_n\right\rbrace $ be the input job set,
    \begin{enumerate}
        \item The job is given \textbf{\textit{online}}, i.e., the algorithm can only learn the job when it is released. But the algorithm does not have to decide the schedule of the job immediately. It can decide the schedule at any time after the job is released and before the job is due.
        \item Each job is characterized with a \textbf{\textit{release time}} $r_j$, a \textbf{\textit{processing time}} $p_j$, and a \textbf{\textit{deadline}} $d_j$(\textit{time-critical task means that the job has a deadline}). Here, we assume $r_j, p_j$ and $d_j$ are all integers. We consider the case that all the jobs have unit processing times, i.e., for any job $j$, there is $p_j=1$.
        \item To ensure a feasible solution, assume that $d_j-r_j\geq p_j+\lambda$ for any $j\in\mathcal{J}$.
    \end{enumerate}
        
    Assume that a job $j$ is scheduled at time $t$, we say that the job $j$ is \textbf{\textit{feasibly scheduled}} if and only if $r_j\leq t$ and $t+p_j\leq d_j$. The objective is to minimize the total number of calibrations while ensuring that all jobs are feasibly scheduled.

    \subsection{Notations and Definitions}
    We use $\mathcal{I}=\left\lbrace I_1,\dots,I_m\right\rbrace $ to denote the set of calibrations, with $t_{I_k}$ being the starting time of calibration $I_k$. Thus, the interval of calibration $I_k$ is $\left[ t_{I_k},t_{I_k}+\lambda+T \right) $, where $\left[ t_{I_k},t_{I_k}+\lambda\right)$ is the calibrating(or activation) interval and $\left[ t_{I_k}+\lambda,t_{I_k}+\lambda+T\right)$ is the calibrated interval. The machine can be used to schedule jobs only during the calibrated interval. The \textbf{\textit{available slot}} at time $t$ means all calibrated machine slot at time $t$, i.e. if there are $c$ calibrations whose calibrated interval include $\left[t,t+1 \right)$, then there are $c$ available slots at time $t$.
        
    The window of a job $j$ is the time interval between the release time and the deadline of the job, i.e. time interval $\left[r_j,d_j\right)$. We classify the jobs into $\alpha$-long and $\alpha$-short window jobs.
    \begin{mydefinition}
        \label{def_alpha_long_and_short}
        A job $j$ is \textbf{\textit{\boldmath{$\alpha$}-long window job}} if $d_j-r_j\geq \alpha T + \lambda$, otherwise the job is a \textbf{\textit{\boldmath{$\alpha$}-short window job}}.
    \end{mydefinition}
        
        
    \subsection{Description of the Solution}
    Since there is an infinite number of machines, the assignment of calibrations to machines is trivial. Thus, a complete solution should include:
    \begin{enumerate}
        \item a set of calibrations, and
        \item an assignment of all jobs to the calibrations.
    \end{enumerate}
    A set of calibrations can be represented by a set of timestamps representing the starting time of each calibration. And a job assignment should specify which job is assigned to which calibration at what time.
        
    However, for the case of unit processing time, we can simplify the structure of the solution by the \textbf{\textit{Earliest Deadline First (EDF)}} algorithm. The algorithm works as follows: The EDF algorithm maintains a dynamic priority queue of current unscheduled jobs ordered by increasing deadline. For each time $t$, assume that there are $r_t$ jobs that are released at time $t$ and there are $m_t$ available calibrated slots at time $t$. The EDF algorithm will first insert the $r_t$ jobs into the priority queue, then remove $m_t$ jobs from the priority queue and schedule these jobs at time $t$. The following lemma guarantees that for any given calibration sets and unit processing time jobs, the EDF algorithm will always find a feasible schedule of the jobs if one exists.
        
    \begin{lem}\cite{DBLP:conf/spaa/BenderBLMP13,DBLP:journals/ior/SimonsS84}
        \label{lemma_offline_EDF}
        Given a calibration set $\mathcal{I}$ and a job set $\mathcal{J}$ with unit processing time jobs, the EDF algorithm finds a feasible schedule of $\mathcal{J}$ on $\mathcal{I}$ if one exists.
    \end{lem}
        
    Lemma \ref{lemma_offline_EDF} implies that given the calibration set as a solution, the EDF algorithm can efficiently construct a feasible job assignment. Thus, the solution of the unit processing time case can be simplified to a calibration set. 
        
    \subsection{Preliminary on Rent Minimization and Machine Minimization}
        
    In the rent minimization problem, the machine can only be used for $T$ time steps after it is activated, while the activation can be instantaneously done. The job is arriving online and is characterized by a release time, a processing time, and a deadline. The objective is to minimize the total number of machines. When $T$ is large enough, rent minimization is degraded to online machine minimization.
        
    Thus, when $\lambda=0$ and $T$ is large enough, our problem degrades to the online machine minimization problem. This implies that our problem shares some common structures with the online machine minimization problem. This fact helps us design our algorithms when the job window is far less than the calibration length $T$. For convenience, we will directly use the online machine minimization algorithm as a subroutine in our algorithm. For the problem of machine minimization with unit processing time jobs, there is a simple way to determine the optimal number of required machines. In the following lemma \ref{lemma_mm_deter} proposed by Kao et al. \cite{DBLP:conf/isaac/KaoCRW12}, the key value involved is the maximum density of jobs. Based on this observation, an $e$-competitive algorithm is proposed for online machine minimization problem \cite{DBLP:journals/corr/DevanurMPY14}, stated in Lemma \ref{lemma_OMM_unit}.
    \begin{mydefinition}[Density and Maximum Density]\cite{DBLP:conf/isaac/KaoCRW12}
        Define the density of $\mathcal{J}$ on $r$ and $d$ as $\rho(\mathcal{J},r,d)$, which is the number of jobs whose release times $\geq r$ and deadlines $\leq d$ dividing the length $d-r$,
        \begin{equation}
            \rho(\mathcal{J},r,d)=\frac{|\left\lbrace j|j\in\mathcal{J},r_j\geq r, d_j\leq d\right\rbrace|}{d-r}.
        \end{equation}
                  
        Define the maximum density of $\mathcal{J}$ as $\rho(\mathcal{J})$, where
        \begin{equation}
            \rho(\mathcal{J})=\max_{0\leq r< d}\left\lbrace \rho(\mathcal{J},r,d)\right\rbrace.
        \end{equation}
    \end{mydefinition}
        
    \begin{lem}\cite{DBLP:conf/isaac/KaoCRW12}
        \label{lemma_mm_deter}
        For a given set $\mathcal{J}$ of unit processing time jobs, let $OPT\left(\mathcal{J}\right)$ be the minimum number of machines that are required to give a feasible schedule of $\mathcal{J}$ and $\rho\left(\mathcal{J}\right)$ is the maximum density of $\mathcal{J}$, then
        \begin{equation}
            OPT\left(\mathcal{J}\right)=\lceil\rho\left(\mathcal{J}\right)\rceil
        \end{equation}
    \end{lem}
        
    \begin{lem}\cite{DBLP:journals/corr/DevanurMPY14}
        \label{lemma_OMM_unit}
        For the problem of online machine minimization with unit processing time jobs, given an input job set $\mathcal{J}$, let ${OPT}$ be the number of machines used in the optimal solution. There exists an online algorithm using at most $\lceil e\cdot {OPT}\rceil$ machines.
    \end{lem}

    \subsection{Lower Bound}
    We propose an adversary argument that obtains a lower bound of $\lambda$. Except that, note that our problem is the generalization of online machine minimization. So, inspired by \cite{DBLP:journals/corr/DevanurMPY14}, a similar construction gives a lower bound of $e$.
    \begin{lem}
        \label{lemma_lower_bound}
        For the problem of online time-critical task scheduling with calibration under unit processing time, no deterministic algorithm can achieve a competitive ratio less than $\max\left\lbrace e,\lambda\right\rbrace$.
    \end{lem}
    \begin{proof}
        First, we prove the lower bound of $\lambda$ for any online algorithm. The interval of a calibration started at time $t$ is $\left[t, t+\lambda+T\right)$. We say the time $\tau$ is covered if and only if there exists a calibration started at $t$ such that $\tau\in\left[t, t+\lambda+T\right)$. Without loss of generality, let $T\gg\lambda$
                  
        For each $\tau\in\left[0,\lambda(\lambda+T)\right)$, if the time $\tau$ is not covered, then the adversary will release $\lambda$ jobs with release times at $\tau$, unit processing times and deadlines at $\tau+\lambda+1$. Thus, the optimal number of calibrations is $1$ (i.e., one calibration at time $\tau-\lambda$) while the algorithm will have to use at least $\lambda$ calibrations. If for every $\tau\in\left[0,\lambda(\lambda+T)\right)$, the time $\tau$ is covered, then by the time of $\lambda(\lambda+T)$, the algorithm has already started at least $\lambda$ calibrations. The adversary can release a job at time $\lambda(\lambda+T)$ with a release time at $\lambda(\lambda+T)$, unit processing time and a deadline at $(\lambda+1)(\lambda+T)$. Then the optimal number of calibrations is $1$ while the algorithm uses at least $\lambda$ calibrations.
                  
        Second, we prove that no online algorithm has a competitive ratio less than $e$. This construction is originally given by Devanur et al. \cite{DBLP:journals/corr/DevanurMPY14}. Assume there exists an $(e-\epsilon)$-competitive algorithm, say \textbf{ALG}, where $\epsilon>0$. Let $\mathcal{J}_t$ be the set of jobs which are released by the time $t$, and let {Online}$(t)$ and Offline$(t)$ be the number of calibrations \textbf{ALG} and the optimal solution for $\mathcal{J}_t$. For each time $t\in\left\lbrace 0,\dots,T-1\right\rbrace$, the adversary will release $\lfloor T^2/(T-t)\rfloor$ jobs, where each job has a deadline of $T+\lambda$. If at some time $\tau$, there is {Online}$(t)>(e-\epsilon)${Offline}$(t)$. then the adversary will stop and this will contradict to the fact that the algorithm \textbf{ALG} is $(e-\epsilon)$-competitive. Now we will prove such $\tau$ must exist.
                  
        Assume that no such $\tau$ exists, that is for every $\tau\in\left\lbrace 0,\dots, T-1\right\rbrace$, there is {Online}$(\tau)\leq(e-\epsilon)${Offline}$(\tau)$. By calculating the maximum density, there is {Offline}$(\tau)\leq\lceil\frac{T^3}{e(T-\tau)(T+\lambda)}\rceil$. So the total number of jobs that \textbf{ALG} scheduled is at most $\sum_{\tau=0}^{T-1}${Online}$(\tau)$:
                  \begin{equation}
                  \begin{aligned}
                  && &\sum_{\tau=0}^{T-1}\text{Online}(\tau)\\
                  &&\leq &(e-\epsilon)\sum_{\tau=0}^{T-1}{\lceil\frac{T^3}{e(T-\tau)(T+\lambda)}\rceil}\\
                  &&\leq&(1-\frac{\epsilon}{e})\sum_{\tau=0}^{T-1}\frac{T^3}{(T-\tau)(T+\lambda)}+\mathcal{O}(T)\\
                  &&=&(1-\frac{\epsilon}{e})\sum_{\tau=0}^{T-1}\left( \frac{T^2}{(T-\tau)}-\frac{\lambda T^2}{(T-\tau)(T+\lambda)}\right)+\mathcal{O}(T)
                  \end{aligned}
                  \end{equation}
                  And the total number of jobs that are released is $\sum_{\tau=0}^{T-1}{\frac{T^2}{T-\tau}}$. So
                  \begin{equation}
                  \begin{aligned}
                  && &\sum_{\tau=0}^{T-1}\text{Online}(\tau)-\sum_{\tau=0}^{T-1}{\frac{T^2}{T-\tau}}\\
                  &&\leq&-\frac{\epsilon}{e}\sum_{\tau=0}^{T-1}\frac{T^2}{T-\tau}-\sum_{\tau=0}^{T-1}\frac{\lambda(e-\epsilon)T^2}{e(T-\tau)(T+\lambda)}+\mathcal{O}(T)\\
                  &&\leq&-\frac{\epsilon}{e}T^2\log T-\frac{\lambda(e-\epsilon)T^2\log T}{e(T+\lambda)}+\mathcal{O}(T)
                  \end{aligned}
                  \end{equation}
                  When $T$ is large enough, the total number of jobs that \textbf{ALG} scheduled is less than the total number of jobs that are released. This is a contradiction. So, such $\tau$ must exist. This concludes the proof.\qed
    \end{proof}

\section{Algorithm for \boldmath{$\alpha$}\unboldmath-Long Window Jobs}
\label{Sec:alpha_long}
        In this section, all results are based on a given input $\mathcal{J}$ where the jobs in $\mathcal{J}$ are all $\alpha$-long window jobs for some specific $\alpha\in\left(0,1\right)$. If the algorithm starts a calibration too early, then the adversary could release some jobs right after the calibration is ended. Thus, the calibration will be wasted since it cannot be used for late-arriving jobs. If the algorithm starts a calibration too late, then the adversary could release some jobs which have the same deadline as the current unscheduled job. This will result in more unscheduled jobs piling up within a small interval (the interval after the calibration is started and before the deadline of the job). Thus, the algorithm will have to do more calibrations in order to finish all jobs. For long window jobs, we find a proper tradeoff that will neither start a calibration too early nor too late.
        \subsection{The Long Jobs Case Algorithm}
        The critical step of our algorithm is that at each time $t$, if the algorithm finds the current calibration set is not enough to schedule all jobs that are due no later than $t+\lambda+T+1$, it will start new calibrations. This condition ensures that the algorithm can delay the jobs so as to group more jobs in one calibration while preventing the job from being delayed too much. This guarantees that the total number of calibrations is still bounded. Repeat this process at time $t$ until all current released jobs with deadline $\leq$ $t+\lambda+T+1$ can be feasibly scheduled.
        
        
        \begin{algorithm}
                  \caption{Algorithm for $\alpha$-Long Window Jobs}
                  \label{algo_long_unit}
                  \begin{algorithmic}[1]
                           \STATE $\mathcal{Q}\gets$ empty priority queue of jobs, ordered by increasing deadline
                           \STATE $\mathcal{I} \gets$ empty set of calibrations
                           \FOR {each time step $t$}
                           \FOR {each job $j$ released at time $t$}
                           \STATE Insert $j$ into $\mathcal{Q}$
                           \ENDFOR
                           \WHILE {There exists a job in $\mathcal{Q}$ with deadline $\leq t+\lambda+T+1$ that cannot be feasibly scheduled if we virtually schedule all jobs of $\mathcal{Q}$ in $\mathcal{I}$ during time interval $\left[t,t+\lambda+T\right]$ by the EDF algorithm}\label{algo_feasibility}
                           \STATE $\mathcal{I}\gets \mathcal{I}\cup \left\lbrace \lceil {\frac{1}{\alpha}} \rceil \times I_{t},I_{t+T}\right\rbrace $\label{algo_calibration}
                           \ENDWHILE
                           \WHILE {$\mathcal{Q}$ is not empty and there are empty calibrated slots at time $t$}
                           \STATE Pop $j$ with the smallest deadline from $\mathcal{Q}$ and schedule $j$ at time $t$\label{algo_true_schedule}
                           \ENDWHILE
                           \ENDFOR
                  \end{algorithmic} 
        \end{algorithm}
        
        Note that the ``Virtually schedule'' in line \ref{algo_feasibility} of Algorithm \ref{algo_long_unit} means the algorithm will not actually schedule any job at this step. The condition in the while loop is only used to decide whether to start several new calibrations or not. The final schedule of any job is always decided in line \ref{algo_true_schedule} of Algorithm \ref{algo_long_unit} which actually mimics the EDF algorithm.
        
        \subsection{The Long Jobs Case Analysis}
        In line \ref{algo_calibration}, we called the $\lceil \frac{1}{\alpha} \rceil$ calibrations at time $t$ and one calibration at time $t+T$ together as \textbf{\textit{a round of calibrations}} starting at time $t$. If this is the $k$-th round of calibrations by Algorithm \ref{algo_long_unit}, let $\mathcal{I}_{k}^{(A)}$ (here $(A)$ is short for ``Algorithm'') be the set of these $(\lceil \frac{1}{\alpha} \rceil+1)$ calibrations and $t_{k}^{(A)}$ be the starting time of $\lceil \frac{1}{\alpha} \rceil$ calibrations. Thus, in $\mathcal{I}_{k}^{(A)}$, there are $\lceil \frac{1}{\alpha} \rceil$ calibrations starting from time $t_{k}^{(A)}$ and one calibration starting from time $t_{k}^{(A)}+T$. Let $\mathcal{I}_{1\sim k}^{(A)}=\mathcal{I}_{1}^{(A)}\cup\dots\cup \mathcal{I}_{k}^{(A)}$, and let $\mathcal{I}_{1\sim0}^{(A)}$ be an empty set. Let $R$ be the number of calibrations that Algorithm \ref{algo_long_unit} will use for $\mathcal{J}$. Let ${OPT}$ be the optimal number of calibrations for $\mathcal{J}$. For convenience, when we say $\mathcal{I}$ is feasible, we mean the whole job set $\mathcal{J}$ can be feasibly scheduled on $\mathcal{I}$.
        
        The feasibility of the Algorithm \ref{algo_long_unit} is obvious by line \ref{algo_feasibility}. We will prove that if all jobs are $\alpha$-long window jobs, then the algorithm has a competitive ratio of  $\lceil \frac{1}{\alpha} \rceil+1$. To do this, we firstly introduce three definitions: the \textbf{\textit{least residual function}}, the \textbf{\textit{compensation set}} and the \textbf{\textit{transition set}}.  
        
        \begin{mydefinition}[Least Residual Function] 
                  \label{def_least_residual_func}
                  Define the least residual function $f_{lr}:\mathbb{N} \rightarrow \left\lbrace 0,1,\dots,OPT \right\rbrace$, 
                  \begin{equation}
                  f_{lr}(k)=\min_{\mathcal{I}: \mathcal{I}\cup\mathcal{I}_{1\sim k}^{(A)}\text{ is feasible.}}{|\mathcal{I}|}.
                  \end{equation}
                  where $f_{lr}(k)$ is the least number of extra calibrations that 
                  makes $\mathcal{I}_{1\sim k}^{(A)}$ feasible.
        \end{mydefinition}
        
        \begin{mydefinition}[Compensation Set]
                  Define the compensation sets $\mathcal{I}_{1}^{(C)},$ $\dots,\mathcal{I}_{R+1}^{(C)}$ as follows. For any $k\in\left\lbrace1,\dots,R+1\right\rbrace$
                  \begin{equation}
                  \mathcal{I}_{k}^{(C)}=\arg \max_{\substack{\mathcal{I}:\mathcal{I}\cup\mathcal{I}_{1\sim k-1}^{(A)}\text{ is feasible }\\\text{and }|\mathcal{I}|=f_{lr}(k-1)}}{\sum_{I\in\mathcal{I}}{t_{I}}}.
                  \end{equation}
        \end{mydefinition}
        
        \begin{mydefinition}[Transition Set]
                  Define the transition sets $\mathcal{I}_{1}^{(T)},$ $\dots,\mathcal{I}_{R+1}^{(T)}$ as follows. For any $k\in\left\lbrace1,\dots,R+1\right\rbrace$
                  \begin{equation}
                  \mathcal{I}_{k}^{(T)}=\mathcal{I}_{1\sim k-1}^{(A)} \cup \mathcal{I}_{k}^{(C)}.
                  \end{equation}
        \end{mydefinition}
        
        After $k-1$ rounds of calibrations, there can be more than one possible calibration set that is feasible to $\mathcal{J}$ and has $f_{lr}(k-1)$ calibrations. We choose the one with maximum total starting time and call it the compensation set $\mathcal{I}_{k}^{(C)}$. Let the $k$-th compensation set together with the first $k-1$ round of calibrations be the $k$-th transition set. By definition, we have $|\mathcal{I}_{1}^{(T)}|=|\mathcal{I}_{1}^{(C)}|=f_{lr}(0)=OPT$, while $f_{lr}(R)=0$,  
$\mathcal{I}_{R+1}^{(C)}=\emptyset$ and $|\mathcal{I}_{R+1}^{(T)}|=(\lceil \frac{1}{\alpha} \rceil+1)R$. Intuitively, the transition set will keep feasible and gradually transforms from the optimal solution to the algorithm solution.  
        
        For a compensation set $\mathcal{I}_{k}^{(C)}$, let $t_{k}^{(C)}$ be the starting time of the calibration in $\mathcal{I}_{k}^{(C)}$ with the minimum starting time. If $\mathcal{I}_{k}^{(C)}$ is an empty set, then let $t_{k}^{(C)}=0$. Thus we have a series of timestamps \boldmath$t_{0}^{(C)},t_{1}^{(C)},\dots,t_{R}^{(C)}$\unboldmath.
        
        We choose the compensation set as the set with maximum total starting time because such set is well-structured. Intuitively, in the beginning period of $\mathcal{I}_{k}^{(C)}$, all calibrations are fully occupied by jobs with small deadlines. Lemma \ref{lemma_opt_structure} shows this structure.
        
        \begin{lem}
                  \label{lemma_opt_structure}
                  For any $k\in\left\lbrace 1,\dots,R\right\rbrace$, use the EDF algorithm to schedule $\mathcal{J}$ on $\mathcal{I}_{k}^{(T)}=\mathcal{I}_{1\sim k-1}^{(A)} \cup \mathcal{I}_{k}^{(C)}$. Let $I_{k}\in\mathcal{I}_{k}^{(C)}$ be the calibration that starts at time $t_{k}^{(C)}$, then there exists $t \in \left( t_{k}^{(C)}+\lambda,t_{k}^{(C)}+\lambda+T \right]$ such that all available time slots in $\left[t_{k}^{(C)}+\lambda,t\right)$ are occupied by jobs, and all these jobs have deadlines $\leq t$. Let $t_{k}^{(F)}$ be the greatest timestamp that satisfies the above property.
        \end{lem}
        \begin{proof}
                  We will prove the lemma by contradiction. Assume that no such $t$ exists. 
                  
                  Let $\mathcal{S}$ be the feasible schedule of $\mathcal{J}$ on $\mathcal{I}^{(T)}_{k}$ by the EDF algorithm. Let $I_k^\prime$ be a calibration started at time $t_{k}^{(C)}+1$ and let $\mathcal{I}_{k}^{(C)\prime}=(\mathcal{I}_{k}^{(C)}\setminus\{I_k\})\cup\{ I_{k}^\prime\}$. We will prove that $\mathcal{J}$ is feasible on $\mathcal{I}_{k}^{(T)\prime}=\mathcal{I}_{1\sim k-1}^{(A)}\cup\mathcal{I}_{k}^{(C)\prime}$. Note that the only difference between $\mathcal{I}_{k}^{(T)\prime}$ and $\mathcal{I}_{k}^{(T)}$ is that the calibration $I_k$ is delayed by one time step. If there exists a feasible schedule of $\mathcal{J}$ on $\mathcal{I}_{k}^{(T)\prime}$, since the number of the calibrations in $\mathcal{I}_{k}^{(C)\prime}$ is the same as that in $\mathcal{I}_{k}^{(C)}$, while the total starting time of $\mathcal{I}_{k}^{(C)\prime}$ is strictly greater than $\mathcal{I}_{k}^{(C)}$, this contradicts the fact that $\mathcal{I}_{k}^{(C)}$ is the one with maximum total starting time.
                  
                  Now, let us prove that if there does not exist such $t$ satisfying the statement of Lemma \ref{lemma_opt_structure}, we have $\mathcal{J}$ is feasible on $\mathcal{I}_{k}^{(T)\prime}$, and this reaches the contradiction. 
                  
                  Firstly, let us consider the case that there is a calibrated slot at time $t_{k}^{(C)}+\lambda$ without a scheduled job.
                  If the calibration $I_{k}$ has a job $j$ scheduled at time $t_{k}^{(C)}+\lambda$ in $\mathcal{S}$, since we have another empty calibrated slot at time $t_{k}^{(C)}+\lambda$, we can move the job $j$ to that empty slot.
                  The other jobs scheduled in $I_{k}$ can also be scheduled in $I_{k}^\prime$ in the same way. Thus, we have a feasible schedule of $\mathcal{J}$ on $\mathcal{I}_{k}^{(T)\prime}$.
                  
                  Secondly, we consider the case that all calibrated slots at time $t_{k}^{(C)}+\lambda$ are occupied. 
                  By our assumption of contradiction, there must be at least one job that is scheduled at time $t_{k}^{(C)}+\lambda$ in $\mathcal{S}$ and has a deadline $>t_{k}^{(C)}+\lambda+1$. Let the one with the greatest deadline be job $j_1$, and let $d_{j_1}$ be the deadline of job $j_1$. Without loss of generality, assume that $j_1$ is scheduled on calibration $I_{k}$ in $\mathcal{S}$. Now reschedule $\mathcal{J}-\left\lbrace j_1\right\rbrace$ on $\mathcal{I}_{k}^{(T)\prime}$ exactly the same as in $\mathcal{S}$. 
                  This assignment of $\mathcal{J}-\left\lbrace j_1\right\rbrace$ is feasible. Let $t\in\left(t_{k}^{(C)}+\lambda,t_{k}^{(C)}+\lambda+T\right]$ be the greatest timestamp such that all available slots between $\left[t_{k}^{(C)}+\lambda,t\right)$ are occupied by jobs. Now we use the following process to assign job $j_1$:
                  \begin{enumerate}
                           \item If $d_{j_1}>t$, schedule job $j_1$ at $t$. This is feasible because there is at least one empty calibrated slot at time $t$. Note that, if $t=t_{k}^{(C)}+\lambda+T$, we also have one empty calibrated slot at time $t$ in $I_{k}^\prime$, while we cannot do so in the calibration $I_{k}$.
                           \item If $d_{j_1}\leq t$, by our assumption of contradiction, during $\left[t_{k}^{(C)}+\lambda,d_{j_1}\right)$, there exists some job with deadline greater than $d_{j_1}$. Let the one with greatest deadline be job $j_2$ and its deadline be $d_{j_2}$. Thus, we know $d_{j_2}>d_{j_1}$. According to the definition of job $j_1$ and $j_2$, we also know that the scheduled time of job $j_2$ is later than the scheduled time of job $j_1$ but earlier than the deadline of job $j_1$. Thus, we can reschedule job $j_1$ at the calibrated slot where $j_2$ is originally scheduled. Repeat the above process for $j_2$.
                  \end{enumerate}
                  
                  \begin{figure}[h]
                           \centering
                           \includegraphics[scale=0.8]{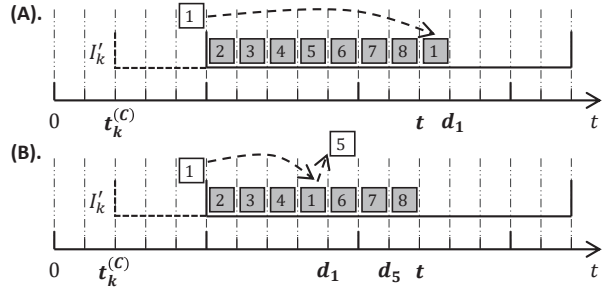}
                           \caption{\boldmath Job $1$ is job $j_1$ and job $5$ is job $j_2$. (A)The case that $d_{j_1}>t$.  (B)The case that $d_{j_1}\leq t$. After the rescheduling, job $j_{1}$ is feasibly scheduled while job $j_{2}$ is pushed out. \unboldmath}
                           \label{Fig0}
                  \end{figure}
                  
                  Figure \ref{Fig0} illustrates the above rescheduling process. The above process will only be executed a finite number of times. Because in each round, we will find some job $j_{k}$ that has a deadline $d_{j_{k}}>d_{j_{k-1}}$. Thus, after finite rounds, we will have $d_{j_{k}}>t$ for some $k$. Finally, we give a feasible schedule of $\mathcal{J}$ on $\mathcal{I}_{k}^{(T)\prime}$.                 \qed
        \end{proof}
        
        By Lemma \ref{lemma_opt_structure}, for every transition set $\mathcal{I}_{k}^{(T)}$, there is a structure related timestamp $t_{k}^{(F)}$. If the compensation set $\mathcal{I}_{k}^{(C)}$ is empty, let $t_{k}^{(F)}=\infty$. Thus we have a series of structure related timestamps \boldmath$t_{0}^{(F)},t_{1}^{(F)},\dots,t_{R}^{(F)}$\unboldmath. There is a connection between $t_{k}^{(C)}$, $t_{k}^{(A)}$ and $t_{k}^{(F)}$, as Lemma \ref{lemma_t_s_a_f} states.
        
        \begin{lem}
                  \label{lemma_t_s_a_f}
                  For every $k\in\left\lbrace 0,1,\dots,R \right\rbrace$, the following inequality holds,
                  \begin{equation}
                  t_{k}^{(C)}-T\leq t_{k}^{(A)}\leq \max\left\lbrace t_{k}^{(C)},t_{k}^{(F)}-\lambda-\alpha T\right\rbrace.
                  \end{equation}
        \end{lem}
        \begin{proof}
                  If $\mathcal{I}_{k}^{(C)}$ is empty, then $t_{k}^{(C)}=0$ and $t_{k}^{(F)}=\infty$, the proposition is true. Now consider the case that $\mathcal{I}_{k}^{(C)}$ is not empty. Let $t_{s}=\max\left\lbrace t_{k}^{(C)},t_{k}^{(F)}-\lambda-\alpha T\right\rbrace$. We will prove by contradiction. Assume that $t_{k}^{(A)}\leq t_{k}^{(C)}-T-1$ or $t_{k}^{(A)}>t_{s}$.
                  
                  \textbf{Case 1}: $t_{k}^{(A)}\leq t_{k}^{(C)}-T-1$.
                  
                  When the algorithm starts a round of calibration at time $t_{k}^{(A)}$, it means that $\mathcal{I}_{1\sim k-1}^{(A)}$ cannot feasibly schedule all jobs with deadline $\leq t_{k}^{(A)}+\lambda+T+1\leq (t_{k}^{(C)}-T-1)+\lambda+T+1=t_{k}^{(C)}+\lambda$. Since $\mathcal{I}_{1\sim k-1}^{(A)}\cup\mathcal{I}_{k}^{(C)}$ is feasible, and all calibration slots in $\mathcal{I}_{k}^{(C)}$ start no earlier than $t_{k}^{(C)}+\lambda$. So all jobs with deadline $\leq t_{k}^{(C)}+\lambda$ must be feasibly scheduled on $\mathcal{I}_{1\sim k-1}^{(A)}$. This contradicts the fact that $\mathcal{I}_{1\sim k-1}^{(A)}$ cannot feasibly schedule all jobs with deadline $\leq t_{k}^{(C)}+\lambda$.
                  
                  \textbf{Case 2}: $t_{k}^{(A)}> t_{s}=\max\left\lbrace t_{k}^{(C)},t_{k}^{(F)}-\lambda-\alpha T\right\rbrace$.
                  
                  We know that the EDF algorithm can feasibly schedule jobs on $\mathcal{I}_{k}^{(T)}$. Let $\mathcal{J}_{f,1}\subseteq \mathcal{J}$ and $\mathcal{J}_{f,2}\subseteq\mathcal{J}$ be the sets of jobs that EDF schedules in $\left[t_{k}^{(C)},t_{k}^{(C)}+\lambda\right)$ and $\left[t_{k}^{(C)}+\lambda,t_{k}^{(F)}\right)$, respectively. Let $\mathcal{J}_{f}=\mathcal{J}_{f,1}\cup\mathcal{J}_{f,2}$. All jobs in $\mathcal{J}_{f,2}$ have deadlines $\leq t_{k}^{(F)}$, thus have release times $\leq t_{k}^{(F)}-\lambda-\alpha T\leq t_{s}$. Now we will prove that all jobs in $\mathcal{J}_{f,1}$ must have release times $\leq t_{s}$. Consider the following two cases:
                  
                  \begin{enumerate}
                           \item If $t_{k}^{(F)}-\lambda-\alpha T<t_{k}^{(C)}$, then by the EDF order, the jobs in $\mathcal{J}_{f,1}$ must have deadlines $\leq t_{k}^{(F)}$, thus have release times $\leq t_s$.
                           \item If $t_{k}^{(C)}\leq t_{k}^{(F)}-\lambda-\alpha T$, then by the EDF order, the jobs in $\mathcal{J}_{f,1}$ with scheduled time $\geq t_{k}^{(F)}-\lambda-\alpha T$ must have deadlines $\leq t_{k}^{(F)}$, so that they have release times $\leq t_s$. And the jobs in $\mathcal{J}_{f,1}$ with scheduled times $<t_{k}^{(F)}-\lambda-\alpha T$ must have release times $<t_{k}^{(F)}-\lambda-\alpha T\leq t_s$.
                  \end{enumerate}
                  So, all jobs in $\mathcal{J}_{f}$ must have release times $\leq \lfloor t_{s}\rfloor \leq t_{k}^{(A)}-1$.
                  
                  Now consider the algorithm at time $t_{k}^{(A)}-1$. At that time, all jobs in $\mathcal{J}_{f}$ have been released. By Lemma \ref{lemma_opt_structure}, all jobs in $\mathcal{J}_{f,2}$ have deadline $\leq t_{k}^{(F)}$, and all slots of $\mathcal{I}_{k}^{(T)}$ in $\left[t_{k}^{(C)}+\lambda,t_{k}^{(F)}\right)$ are occupied by jobs if we run the EDF algorithm. 
                  
                  Thus, we have the number of slots of $\mathcal{I}_{1\sim k-1}^{(A)}$ in $\left[t_{k}^{(C)}+\lambda,t_{k}^{(F)}\right)$ is not enough, which means the jobs in $\mathcal{J}_{f}$ cannot be feasibly scheduled on $\mathcal{I}_{1\sim k-1}^{(A)}$. According to Algorithm \ref{algo_long_unit} at line \ref{algo_feasibility}, the algorithm will start a new round of calibration at time $t_{k}^{(A)}-1$. This contradicts the fact that the next round of calibration is at time $t_{k}^{(A)}$.
                  \qed
        \end{proof}
        
        Based on the connection stated in Lemma \ref{lemma_t_s_a_f}, we now prove that $f_{lr}$ is strictly decreasing until the value of the function hits $0$.
        \begin{lem}
                  \label{lemma_flr_nondecrease}
                For any $k\in\mathbb{N}^{+}$, the following inequality holds,
                  \begin{equation}
                  f_{lr}(k)\leq \max\left\lbrace 0,f_{lr}(k-1)-1\right\rbrace.
                  \end{equation}
        \end{lem}
        \begin{proof}
                  By Definition \ref{def_least_residual_func}, the function $f_{lr}$ must be non-increasing. So, if $f_{lr}(x)=0$, then for any $k> x$, $f_{lr}(k)=0$. Thus, we only consider the case where $f_{lr}(k-1)>0$, and show $f_{lr}(k)\leq f_{lr}(k-1)-1$.
                  
                  Let $I_{k}$ be the calibration with the least starting time in $\mathcal{I}_{k}^{(C)}$. Consider a new calibration set $\mathcal{I}_{new}=\mathcal{I}_{1\sim k}^{(A)}\cup(\mathcal{I}_{k}^{(C)}\setminus \left\lbrace I_{k} \right\rbrace)$. We will prove that $\mathcal{I}_{new}$ can feasibly schedule $\mathcal{J}$, thus $f_{lr}(k)\leq| \mathcal{I}_{k}^{(C)}\setminus\left\lbrace I_{k} \right\rbrace| \leq f_{lr}(k-1)-1$.
                  
                  By Lemma \ref{lemma_t_s_a_f}, we have $t_{k}^{(C)}-T\leq t_{k}^{(A)} \leq \max\left\lbrace t_{k}^{(C)},t_{k}^{(F)}-\lambda-\alpha T\right\rbrace$. Consider the following two cases:
                  
                  \textbf{Case 1}: $t_{k}^{(C)}-T\leq t_{k}^{(A)}\leq t_{k}^{(C)}$.
                  
                  In this case, $I_{k}^{(A)}$ includes $\lceil \frac{1}{\alpha}\rceil$ calibrations at time $t_{k}^{(A)}$ and one calibration at time $t_{k}^{(A)}+T$. Since $\left[t_{k}^{(C)}+\lambda,t_{k}^{(C)}+\lambda+T\right)\subseteq \left[ t_{k}^{(A)}+\lambda,t_{k}^{(A)}+\lambda+2T\right)$, the calibration $I_{k}$ is covered by $I_{k}^{(A)}$. Since the EDF algorithm can feasibly schedule $\mathcal{J}$ on $\mathcal{I}_{1\sim k-1}^{(A)}\cup\mathcal{I}_{k}^{(C)}$, we can simply move the assignment of jobs in $I_{k}$ to $I_{k}^{(A)}$. Thus $\mathcal{I}_{new}$ can feasibly schedule $\mathcal{J}$.
                  
                  \textbf{Case 2}: $t_{k}^{(C)}<t_{k}^{(A)}\leq 
t_{k}^{(F)}-\lambda-\alpha T $. 
                  
                  The EDF algorithm can feasibly schedule $\mathcal{J}$ on $\mathcal{I}_{k}^{(T)}$. Let $\mathcal{J}_{s}$ be the set of jobs that are scheduled in time interval $\left[t_{k}^{(C)}+\lambda,t_{k}^{(F)}\right)$.
                  
                  Let $I_{k}^{(A)}\in\mathcal{I}_{k}^{(A)}$ be a calibration that starts at time $t_{k}^{(A)}$.  Consider the calibration set $\mathcal{I}_{tmp}=\mathcal{I}_{1\sim k-1}^{(A)}
                  \cup( \mathcal{I}_{k}^{(C)}\setminus \{I_{k}\} )
                  \cup\{ I_{k}^{(A)} \}$. For each time $t\in\left[t_{k}^{(C)}+\lambda,t_{k}^{(A)}+\lambda\right)$, $\mathcal{I}_{tmp}$ contains exactly one calibrated slot fewer than $\mathcal{I}_{k}^{(T)}$. 
                  
                  Now, let us run the EDF algorithm to schedule $\mathcal{J}_s$ on $\mathcal{I}_{tmp}$ during time interval $\left[t_{k}^{(C)}+\lambda,t_{k}^{(F)}\right)$. Since during this time interval, all calibration slots are occupied by jobs if the calibration set is $\mathcal{I}_{k}^{(T)}$, and the calibration set $\mathcal{I}_{tmp}$ contains $(t_{k}^{(A)}-t_{k}^{(C)})$ calibrated slots fewer than $\mathcal{I}_{k}^{(T)}$, the EDF algorithm will fail to schedule exactly $(t_{k}^{(A)}-t_{k}^{(C)})$ jobs. Let these jobs be $\mathcal{J}_{f}$.
                  
                  
                  The job assignment of $\mathcal{J}$ on $\mathcal{I}_{new}$ works as follows where $\mathcal{I}_{new}=\mathcal{I}_{tmp}\cup(\mathcal{I}_{k}^{(A)}\setminus \{I_{k}^{(A)}\})$:
                  
                  \begin{enumerate}
                           \item Use the EDF algorithm to schedule $\mathcal{J}-\mathcal{J}_{f}$ on $\mathcal{I}_{tmp}$.
                           \item Use the EDF algorithm to schedule $\mathcal{J}_{f}$ on $\mathcal{I}_{k}^{(A)}-\left\lbrace I_{k}^{(A)}\right\rbrace$.
                  \end{enumerate}
                  The reassigning process is illustrated in Figure \ref{Fig1}. It is trivial that $\mathcal{J}-\mathcal{J}_{f}$ is feasible on $\mathcal{I}_{tmp}$. We only have to prove that $\mathcal{J}_{f}$ can be feasibly scheduled on $\mathcal{I}_{k}^{(A)}-\left\lbrace I_{k}^{(A)}\right\rbrace$. In fact, we will only use the $\lceil\frac{1}{\alpha}\rceil-1$ calibrations at time $t_{k}^{(A)}$.
                  
                  \begin{figure}[h]
                           \centering
                           \includegraphics[scale=0.8]{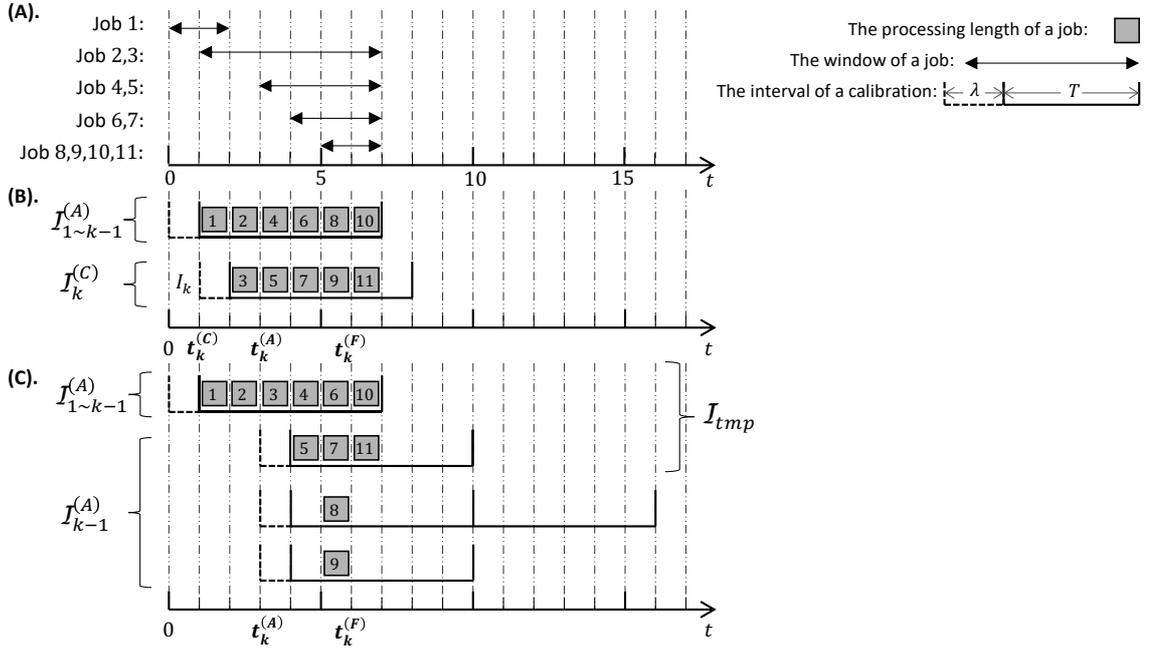}
                           \caption{\boldmath(A)The diagram shows the window of the job 1 to job 11. (B)The diagram shows a feasible schedule of $\mathcal{J}$ on $\mathcal{I}^{(T)}_{k}$, and $\mathcal{J}_{s}=$ $\left\lbrace2,3,4,5,6,7,8,9\right\rbrace$. (C)The diagram shows the reassigned $\mathcal{J}$ on $\mathcal{I}_{new}$. The set $\mathcal{J}_{f}$ is $\left\lbrace 8,9 \right\rbrace$ in this example.\unboldmath}
                           \label{Fig1}
                  \end{figure}
                  
                  We will first prove an observation that no job $j$ in $\mathcal{J}_{f}$ satisfies that $r_j<t_{k}^{(A)}+\lambda$ and $d_j< t_{k}^{(A)}+\lambda+\alpha T$. Assume that such a job $j$ exists. Let $\mathcal{J}_{s}=\mathcal{J}_{1}\cup\mathcal{J}_{2}$, where $\mathcal{J}_{1}$ is the set of jobs that have deadlines $\leq d_{j}$ and $\mathcal{J}_{2}$ is the set of jobs that have deadlines $>d_{j}$. Since all jobs are $\alpha$-long jobs, the release times of jobs in $\mathcal{J}_{1}$ must be $\leq d_{j}-\lambda-\alpha T< t_{k}^{(A)}$. This implies all jobs in $\mathcal{J}_{1}$ must be released no later than $t_{k}^{(A)}-1$. Consider the time at $t_{k}^{(A)}-1$, the algorithm will not start a new round of calibrations, 
                 which implies that $\mathcal{J}_{1}$ can be feasibly scheduled on $\mathcal{I}_{1\sim k-1}^{(A)}$. So use the EDF algorithm to schedule $\mathcal{J}_{s}$ on $\mathcal{I}_{tmp}$, $j\in\mathcal{J}_{1}$ must be feasibly scheduled. Job $j$ cannot appear in $\mathcal{J}_{f}$. This is a contradiction.
                  
                  We now prove that $\mathcal{J}_{f}$ can be feasibly scheduled on the $\lceil\frac{1}{\alpha}\rceil-1$ calibrations at time $t_{k}^{(A)}$. Construct a new job set $\mathcal{J}_{f}'$ which delays the release time of all jobs in $\mathcal{J}_{f}$ after $t_{k}^{(A)}$. The formal construction is as follows: For any job $j\in\mathcal{J}_{f}$,
                  \begin{enumerate}
                           \item If $r_{j}\geq t_{k}^{(A)}+\lambda$, add $j$ to $\mathcal{J}_{f}'$.
                           \item If $r_{j}<t_{k}^{(A)}+\lambda$, by the above assertion, $d_{j}\geq t_{k}^{(A)}+\lambda+\alpha T$. So let job $j'$ with $r_{j}'=t_{k}^{(A)}+\lambda$ and $d_{j}'=d_{j}$. Add $j'$ to $\mathcal{J}_{f}'$.
                  \end{enumerate}
                  Since all jobs in $\mathcal{J}_{f}'$ have release times $\geq t_{k}^{(A)}+\lambda$ and deadlines $\leq t_{k}^{(F)}\leq t_{k}^{(C)}+\lambda+T< t_{k}^{(A)}+\lambda+T$, schedule $\mathcal{J}_{f}'$ on $\lceil\frac{1}{\alpha}\rceil-1$ calibrations is equivalent to schedule $\mathcal{J}_{f}'$ on $\lceil\frac{1}{\alpha}\rceil-1$ machines. Consider the maximum density of $\mathcal{J}_{f}'$,
                  \begin{equation}
                  \begin{aligned}
                  &&\rho(\mathcal{J}_{f}')\leq&\frac{|\mathcal{J}_{f}'|}{\alpha T}
                  =\frac{t_{k}^{(A)}-t_{k}^{(C)}}{\alpha T}
                  \leq\frac{t_{k}^{(F)}-\lambda-\alpha T-t_{k}^{(C)}}{\alpha T}\\
                  &&\leq&\frac{T-\alpha T}{\alpha T}
                  \leq\lceil\frac{1}{\alpha}\rceil-1
                  \end{aligned}
                  \end{equation}
                  By Lemma \ref{lemma_mm_deter}, $\mathcal{J}_{f}'$ can be feasibly scheduled on $\lceil \frac{1}{\alpha} \rceil-1$ machines. Thus $\mathcal{J}_{f}$ is feasible, too. This concludes the proof.
                  \qed
        \end{proof}
        
        With the above property of $f_{lr}$, we have the following theorem.
        \begin{thm}
                  \label{theorem_unit_long_competitive}
                  Algorithm \ref{algo_long_unit} has a competitive ratio of $\lceil \frac{1}{\alpha} \rceil+1$.
        \end{thm}
        \begin{proof}
                  It is obvious that $f_{lr}(0)=OPT$, and by Lemma \ref{lemma_flr_nondecrease}, $f_{lr}(OPT)$ must be $0$. This means that Algorithm \ref{algo_long_unit} will start at most $OPT$ rounds of calibrations. So the final cost of Algorithm \ref{algo_long_unit} is at most $\left(\lceil \frac{1}{\alpha}\rceil+1\right)\cdot R\leq 
\left(\lceil \frac{1}{\alpha}\rceil+1\right)\cdot {OPT}$. This concludes the proof.
                  \qed
        \end{proof}

\section{Algorithm for \boldmath{$\alpha$}\unboldmath-Short Window Jobs}
\label{Sec:alpha_short}
        In this section, all results are based on a given input $\mathcal{J}$, and the jobs in $\mathcal{J}$ are all $\alpha$-short window jobs, where $\alpha\in\left(0,1\right)$.
        \subsection{From General $\lambda\geq 0$ to $\lambda=0$}
        We will first prove that the general case of any integer $\lambda\geq 0$ can be solved by the algorithm for the special case of $\lambda=0$ with only $\mathcal{O}(\lambda)$ loss in the factor of the competitive ratio. Firstly, given an input job set $\mathcal{J}$ for $\lambda\geq 0$, we will construct another job set $\mathcal{J}_{\lambda=0}$. The construction works as follows.
        \begin{itemize}
                  \item For any job $j\in\mathcal{J}$, we have $d_{j}-r_{j}\geq \lambda+1$ and $p_{j}=1$. Let $j^{\prime}$ be a new job with $r_{j^{\prime}}=r_{j}$, $p_{j^{\prime}}=1$ and $d_{j^{\prime}}=d_{j}-\lambda$. Let $j^{\prime}$ be a job in $\mathcal{J}_{\lambda=0}$.
        \end{itemize}
        
        \begin{lem}
                  \label{lemma_lambda=0_and_lambda}
                  Let the optimal number of calibrations for $\mathcal{J}$ and $\mathcal{J}_{\lambda=0}$ be $OPT$ and ${OPT}_{\lambda=0}$, respectively. Then $OPT_{\lambda=0}\leq (\lambda+1)OPT$.
        \end{lem}
        \begin{proof}
                  For every calibration $I$ in the solution of $\mathcal{J}$, let the calibrated interval of $I$ be $\left[t, t+T\right)$. Let $I_{0},\dots,I_{\lambda-1}$ be calibrations with calibrated interval $\left[t-\lambda, t-\lambda+T\right)$, and let $I_{\lambda}$ be a calibration with calibrated interval $\left[t,t+T\right)$. Let $j_1,\dots,j_m \in\mathcal{J}$ be jobs scheduled on $I$, we will schedule $j_{1}^{\prime},\dots, j_{m}^{\prime}\in\mathcal{J}_{\lambda=0}$ as follows.
                  \begin{enumerate}
                           \item If the scheduled time of $j_{x}$(that is $t_{j_{x}}$) satisfied $t_{j_x}+\lambda<d_{j_x}$, then schedule job $j_{x}^{\prime}$ at time $t_{j_x}$ in calibration $I_{\lambda}$.
                           \item If the scheduled time of $j_{x}$(that is $t_{j_{x}}$) satisfied $t_{j_x}+\lambda\geq d_{j_x}$, then schedule job $j_{x}^{\prime}$ at time $d_{j_x}-\lambda-1(\geq t_{j_x}+1-\lambda-1=t_{j_x}-\lambda)$ in calibration $I_{x\%\lambda}$.
                  \end{enumerate}
                  We will prove that the schedule of $\mathcal{J}_{\lambda=0}$ is feasible. To do this, all we have to prove is that no two jobs in $I_{k}$($k\neq\lambda$) are scheduled at the same time slot. Assume that $j_{x}^{\prime}$ and $j_{y}^{\prime}$ are scheduled in $I_{k}$ at time $\tau$. This implies that the scheduled time of $j_x$ and $j_y$ satisfied $t_{j_x}, t_{j_y}\in\left[\tau+1,\tau+\lambda\right]$. And by the above assigning rules, $t_{j_y}-t_{j_x}\geq \lambda$, which is a contradiction. So the schedule of $\mathcal{J}_{\lambda}$ is feasible, thus $OPT_{\lambda=0}\leq (\lambda+1)OPT$.
                  \qed
        \end{proof}
        
        Secondly, we will show that given the solution of $\mathcal{J}_{\lambda}$, how to obtain a feasible solution of $\mathcal{J}$:
        \begin{enumerate}
                  \item For a calibration $I^{\prime}$ with calibrated interval $\left[t,t+T\right)$ for $\mathcal{J}$. Let $I$ with calibrating interval $\left[t,t+\lambda\right)$ and calibrated interval $\left[t+\lambda,t+\lambda+T\right)$ be a calibration of the solution for $\mathcal{J}$.
                  \item For a job $j^{\prime}\in\mathcal{J}_{\lambda}$ scheduled in $I^{\prime}$ at time $\tau$. Schedule job $j\in\mathcal{J}$ at time $\tau+\lambda(<d_{j^\prime}+\lambda=d_{j})$ in $I$.
        \end{enumerate}
        \subsection{Algorithm for the Case of $\lambda=0$}
        As for the case of $\lambda=0$, motivated by the work of Fineman and Sheridan \cite{DBLP:conf/spaa/FinemanS15}, we find that this special case is closely related to the problem of online machine minimization. The intuition is that: If the window of all jobs is within $\left[t,t+T\right)$ for some $t$, then we can transform the algorithm of machine minimization into the algorithm of calibration minimization as follows: When the algorithm of machine minimization opens up a machine at time $t$, our algorithm of calibration minimization will start a calibration at time $t$.
        
        Further, we can divide the jobs into many non-intersected sets, and the jobs in each subset of $\mathcal{J}$ fall within $\left[t,t+T\right)$ for some $t$. More specifically, we divide $\mathcal{J}$ by the starting time of the jobs. Let $\mathcal{J}_{k}\subseteq \mathcal{J}$ be the set of jobs that the release time of any job in $\mathcal{J}_{k}$ falls within $\left[k(T-\lfloor\alpha T \rfloor),(k+1)(T-\lfloor\alpha T \rfloor) \right)$, for all $k\in\mathbb{N}$. It is trivial that $\cup_{k\in\mathbb{N}}{\mathcal{J}_{k}}$ will include every job in $\mathcal{J}$ and no two $\mathcal{J}_{k}$ with different index $k$ intersect with each other. Since the input jobs are all $\alpha$-short window jobs, the window of each job in a given $\mathcal{J}_{k}$ falls within $\left[k( T-\lfloor\alpha T \rfloor),k(T-\lfloor\alpha T \rfloor)+T \right)$, an exactly $T$-length long interval. Thus, we can use the online machine minimization algorithm for each $\mathcal{J}_{k}$ for $k\in\mathbb{N}$. Let $\mathcal{I}_{k}$ be the set of calibrations started for $\mathcal{J}_{k}$, $k\in\mathbb{N}$. Refer to Algorithm \ref{algo_short_unit_deter} for the detailed description.
        
        For simplicity, we do not explicitly schedule jobs on calibrations in Algorithm \ref{algo_short_unit_deter}. However, by Lemma \ref{lemma_offline_EDF}, the algorithm can use the EDF algorithm to schedule each $\mathcal{J}_{k}$ in $\mathcal{I}_{k}$ under online style. This will output a feasible schedule of $\mathcal{J}$.
        \begin{algorithm}
                  \caption{Deterministic Algorithm for $\alpha$-Short Window Jobs}
                  \label{algo_short_unit_deter}
                  \begin{algorithmic}[1]
                           \FOR {each time step $t$}
                           \FOR {each job $j$ released at time $t$}
                           \STATE Let job $j$ be in the set $\mathcal{J}_{\lfloor t/{\left(T-\lfloor\alpha T \rfloor\right)} \rfloor}$
                           \ENDFOR
                           \FOR {each non-empty $\mathcal{J}_{k}$}
                           \STATE Using online machine minimization algorithm as a black box for $\mathcal{J}_{k}$ at time $t$
                           \IF {the black box algorithm opens up $m$ machine}
                           \STATE $\mathcal{I}_{k}=\mathcal{I}_{k}\cup\left\lbrace m\times I_{t}\right\rbrace$
                           \ENDIF
                          \ENDFOR
                           \ENDFOR
                  \end{algorithmic} 
        \end{algorithm}
        \subsection{The Short Jobs Case Analysis}
        The feasibility of Algorithm \ref{algo_short_unit_deter} is obvious. We will prove the competitive ratio of Algorithm \ref{algo_short_unit_deter} is $(e+1)\lceil \frac{2}{1-\alpha}\rceil$. 
        
        \begin{lem}
                  \label{lemma_unit_short_deter_competitive}
                  The algorithm \ref{algo_short_unit_deter} has a competitive ratio of $ (e+1)\lceil \frac{2}{1-\alpha}\rceil$.
        \end{lem}
        \begin{proof}
                  First, we introduce some notations. Let ${alg}_{k}$ be the number of calibrations that Algorithm \ref{algo_short_unit_deter} started for the job set $\mathcal{J}_{k}$, $k\in\mathbb{N}$. Let ${opt}_{k}^{(m)}$ be the optimal number of machines that the online machine minimization used for $\mathcal{J}_{k}$, $k\in\mathbb{N}$. Consider any optimal solution of online calibration scheduling problem of $\mathcal{J}$, let ${opt}_{\tau}$ be the number of calibrations that the optimal solution started at time $\tau$, $\tau\in\mathbb{N}$. For the simplicity of later proof, we define ${opt}_{\tau}=0$ for any integer $\tau<0$.
                  
                  In the optimal solution, let all these calibrations that intersect the time interval $\left[k(T-\lfloor \alpha T\rfloor),k(T-\lfloor \alpha T\rfloor)+T \right)$ together be $\mathcal{I}_{k}^{(opt)}$. Define $f(k)=|\mathcal{I}_{k}^{(opt)}|$ as the number of calibrations in $\mathcal{I}_{k}^{(opt)}$. Let $C$ be the cost of Algorithm \ref{algo_short_unit_deter} for $\mathcal{J}$, by Lemma \ref{lemma_OMM_unit}, 
                  \begin{equation}
                  \label{eq_competitive_analysis_short_deter}
                  \begin{aligned}
                  &&C=&\sum_{k\geq 0}{{alg}_{k}}
                  \leq\sum_{k\geq 0}{\lceil e\cdot {opt}_{k}^{(m)}\rceil}\\
                  &&\leq&\sum_{k\geq 0} e\cdot|\mathcal{I}_{k}^{(opt)}|
                  + \sum_{k\geq 0}\textbf{1}\left(|\mathcal{I}_{k}^{(opt)}|> 0\right)\\
                  &&\leq&\lceil \frac{2}{1-\alpha}\rceil\cdot (e+1)\cdot\sum_{\tau}{{opt}_{\tau}}\\
                  &&=&\lceil \frac{2}{1-\alpha}\rceil\cdot (e+1)\cdot {OPT}.
                  \end{aligned}
                  \end{equation}
                  
                  The second inequality in Equation (\ref{eq_competitive_analysis_short_deter}) comes from the fact that $\mathcal{J}_{k}$ can be feasibly scheduled on $\mathcal{I}_{k}^{(opt)}$. So the number of machines used for $\mathcal{J}_{k}$ must be no more than the number of calibrations in $\mathcal{I}_{k}^{(opt)}$. So,
                  \begin{equation}
                  {opt}_{k}^{(m)}\leq |\mathcal{I}_{k}^{(opt)}|.
                  \end{equation}
                  
                  The third inequality in Equation 
                  ($\ref{eq_competitive_analysis_short_deter}$) comes from two facts:
                  
                  First, note that $|\mathcal{I}_{k}^{(opt)}|=\sum_{\tau=k T-k\lfloor \alpha T\rfloor-T+1}^{(k+1) T-k\lfloor \alpha T\rfloor-1}{{opt}_{\tau}}$. Consider a simple counting of $k$. That is given any $\tau$, count the number of integers $k$ satisfying 
                  \begin{equation}
                  \tau\geq k T-k\lfloor\alpha T\rfloor-T+1\text{ and }\tau< (k+1) T-k\lfloor \alpha T\rfloor.
                  \end{equation}
                  This implies
                  \begin{equation}
                  k>\frac{\tau-T}{T-\lfloor\alpha T\rfloor}\text{ and }k\leq\frac{\tau+T-1}{T-\lfloor\alpha T\rfloor}.
                  \end{equation}
                  And the number of such $k$ will be bounded by
                  \begin{equation}
                  \begin{aligned}
                  && &\lceil\frac{\tau+T-1}{ T-\lfloor\alpha T\rfloor}-\frac{\tau-T}{T-\lfloor\alpha T\rfloor}\rceil
                  \leq\lceil\frac{2}{1-\alpha}\rceil.
                  \end{aligned}
                  \end{equation}
                  Second, a calibration in the optimal solution can lead to at most $\lceil\frac{2}{1-\alpha}\rceil$ such $\textbf{1}\left(|\mathcal{I}_{k}^{(opt)}|> 0\right)$ to be non-zero. Thus $\sum_{k\geq 0}\textbf{1}\left(|\mathcal{I}_{k}^{(opt)}|> 0\right)$ will be bounded by $\lceil\frac{2}{1-\alpha}\rceil\cdot {OPT}$. This concludes the proof.
                  \qed
        \end{proof}
        
        Combining the above results together, we obtained an algorithm for the $\alpha$-short window jobs with integer $\lambda\geq 0$.
        \begin{thm}
                  \label{theorem_unit_short_jobs}
                  For the online time-critical task scheduling with calibration problem when all jobs are $\alpha$-short window jobs and the calibrating interval length is integer $\lambda\geq 0$, there exists an algorithm for this problem with a competitive ratio of $(e+1)(\lambda+1)\lceil \frac{2}{1-\alpha}\rceil$
        \end{thm}
        \begin{proof}
                  It follows directly from Lemma \ref{lemma_lambda=0_and_lambda} and Lemma \ref{lemma_unit_short_deter_competitive}.
                  \qed
        \end{proof}

\section{Integrated Algorithm}
\label{Sec:Integrated Algorithm}
        
        Together, we can develop the algorithm for unit processing time case. By choosing a proper value of $\alpha$, and let any input job $j$ be either in the subroutine of long job algorithm or short job algorithm depending on whether the job $j$ is $\alpha$-long window or $\alpha$-short window. The integration process is illustrated in Algorithm \ref{algo_integrated} in pseudo-code. Note that in the pseudo-code,  both the subroutine $\mathcal{R}_{long}$ and $\mathcal{R}_{short}$ are online algorithms, thus they will receive inputs at each time steps $t$. When we say ``feed a job $j$ to the subroutine at time step $t$'', it means the subroutine will receive an input job $j$ at time step $t$. Theorem \ref{theorem_unit} gives the upper bound of Algorithm \ref{algo_integrated}.
        
        \begin{algorithm}
                  \caption{Integrated Algorithm for Unit Processing Time Jobs}
                  \label{algo_integrated}
                  \begin{algorithmic}[1]
                    \STATE $\mathcal{R}_{long} \gets$ Algorithm \ref{algo_long_unit} as a subroutine
                    \STATE $\mathcal{R}_{short} \gets$ Algorithm \ref{algo_short_unit_deter} as a subroutine
                    \FOR {each time step $t$}
                    \FOR {each job $j$ released at time $t$}
                           
                        \IF {$d_j-r_j \geq \alpha T + \lambda$}
                            \STATE Feed $j$ to $\mathcal{R}_{long}$
                        \ELSE
                            \STATE Feed $j$ to $\mathcal{R}_{short}$
                        \ENDIF
                    \ENDFOR
                    \ENDFOR
                    \STATE Combine the output of $\mathcal{R}_{long}$ and $\mathcal{R}_{short}$ together as a complete solution
                  \end{algorithmic} 
        \end{algorithm}

        \begin{thm}
                  \label{theorem_unit}
                  There exists a $3(e+1)\lambda+3e+7$-competitive deterministic algorithm for online calibration scheduling problem with unit processing time jobs. For the special case that $\lambda=0$, there exists a $3e+7(\approx 15.16)$-competitive deterministic algorithm.
        \end{thm}
        \begin{proof}
                  \sloppy
                  By Theorem \ref{theorem_unit_long_competitive} and Theorem \ref{theorem_unit_short_jobs}, we know there exists an $(e+1)(\lambda+1)\lceil \frac{2}{1-\alpha}\rceil+\lceil\frac{1}{\alpha}\rceil+1$-competitive deterministic algorithm for $\alpha\in\left(0,1\right)$. The results are achieved when $\alpha=\frac{1}{3}$.
                  \qed
        \end{proof}

\section{Conclusion}
\label{Sec:Conclusion}
        In this paper, we study online scheduling with calibration while calibrating a machine will require certain time units. We give an asymptotically optimal algorithm for this problem when all the jobs have unit processing time. And for the special case that calibrating a machine is instantaneous, our problem degrades to rent minimization problem, and our algorithm achieves a better competitive ratio than previous results.
        
        However, the current upper and lower bound gap is still large for the special cases of unit processing time and $\lambda=0$. One open problem is to narrow these gaps. Another interesting problem is to consider the number of machines we use in these algorithms. When the number of calibrations we use is asymptotically optimal, can we also guarantee that the number of machines we use is also asymptotically optimal?



\bibliographystyle{elsarticle-num} 
\bibliography{mybibfile.bib}






\end{document}